\documentclass[11pt,preprint]{aastex}

\usepackage{natbib}
\begin{document}

\title{Constraints on the origin of cosmic rays above $10^{18}$~eV from large scale 
anisotropy searches in data of the Pierre Auger Observatory}

\author{\textbf{The Pierre Auger Collaboration$^{\dag}$}\\
P.~Abreu$^{63}$, M.~Aglietta$^{51}$, M.~Ahlers$^{94}$, E.J.~Ahn$^{81}$, I.F.M.~Albuquerque$^{15}$, D.~Allard$^{29}$, I.~Allekotte$^{1}$, J.~Allen$^{85}$, P.~Allison$^{87}$, A.~Almela$^{11,\: 7}$
% ...
\\ }
\affil{
$^{\dag}$ Av. San Mart\'in Norte 306, 5613 Malarg\"ue, Mendoza, Argentina; www.auger.org \\
$^{1}$ Centro At\'{o}mico Bariloche and Instituto Balseiro (CNEA-UNCuyo-CONICET), San 
Carlos de Bariloche, 
Argentina \\
% ...
$^{7}$ Instituto de Tecnolog\'{\i}as en Detecci\'{o}n y Astropart\'{\i}culas (CNEA, CONICET, UNSAM), 
Buenos Aires, 
Argentina \\
% ...
$^{11}$ Universidad Tecnol\'{o}gica Nacional - Facultad Regional Buenos Aires, Buenos Aires,
Argentina \\
% ...
$^{15}$ Universidade de S\~{a}o Paulo, Instituto de F\'{\i}sica, S\~{a}o Paulo, SP, 
Brazil \\
% ...
$^{29}$ Laboratoire AstroParticule et Cosmologie (APC), Universit\'{e} Paris 7, CNRS-IN2P3, 
Paris, 
France \\
% ...
$^{51}$ Istituto di Fisica dello Spazio Interplanetario (INAF), Universit\`{a} di Torino and 
Sezione INFN, Torino, 
Italy \\
% ...
$^{63}$ LIP and Instituto Superior T\'{e}cnico, Technical University of Lisbon, 
Portugal \\
% ...
$^{81}$ Fermilab, Batavia, IL, 
USA \\
% ...
$^{85}$ New York University, New York, NY, 
USA \\
% ...
$^{87}$ Ohio State University, Columbus, OH, 
USA \\
% ...
$^{94}$ University of Wisconsin, Madison, WI, 
USA \\
% ...
(\ddag) Deceased \\
(a) at Konan University, Kobe, Japan \\
(b) now at the Universidad Autonoma de Chiapas on leave of absence from Cinvestav \\
(f) now at University of Maryland \\
(h) now at NYU Abu Dhabi \\
(i) now at Universit\'{e} de Lausanne \\
}

\begin{abstract}
A thorough search for large scale anisotropies in the distribution of arrival directions 
of cosmic rays detected above $10^{18}$~eV at the Pierre Auger Observatory is 
reported. For the first time, these large scale anisotropy searches are performed
as a function of both the right ascension and the declination and expressed 
in terms of dipole and quadrupole moments. 
Within the systematic uncertainties, no significant deviation from isotropy is revealed. 
Upper limits on dipole and quadrupole amplitudes are derived under the hypothesis
that any cosmic ray anisotropy is
dominated by such moments in this energy range. 
These upper limits provide constraints on the production of cosmic rays above $10^{18}$~eV,
since they allow us to challenge an origin from stationary galactic sources densely 
distributed in the galactic disk and emitting predominantly light particles in all directions.
\end{abstract}


\newpage



The large scale distribution of arrival directions of Ultra-High Energy Cosmic Rays 
(UHECRs) as a function of the energy is a key observable to provide further understanding 
of their origin. Above $\simeq 0.25$~EeV, the most stringent 
bounds ever obtained on the dipole component in the equatorial 
plane were recently reported, being 
below 2\% at 99\% $C.L.$ for EeV energies~\citep{AugerAPP2011}. Such a sensitivity provides
some constraints upon scenarios in which dipolar anisotropies could be imprinted in the distribution 
of arrival directions as the result of the escape of UHECRs from the Galaxy up to the ankle
energy~\citep{Ptuskin1993,Candia2003,Giacinti2011}. 
On the other hand, if UHECRs above 1~EeV have already a predominant extragalactic 
origin~\citep{Hillas1967,Blumenthal1970,Berezinsky2006,Berezinsky2004}, their angular 
distribution is expected to be isotropic to a high level. Thus, the study of large scale
anisotropies at EeV energies would help in establishing whether the origin of UHECRs is
galactic or extragalactic in this energy range. 

The upper limits aforementioned are based on first harmonic analyses of the right ascension
distributions in several energy ranges. The analyses benefit from the almost uniform
directional exposure in right ascension of any ground based observatory operating with 
high duty cycle, but are not sensitive to a dipole component along the Earth
rotation axis. In contrast, using the large amount of data collected by the surface
detector array of the Pierre Auger Observatory, we report in this letter on searches
for dipole and quadrupole patterns significantly standing out above the background noise
whose components are functions of \textit{both} the right ascension and the declination (a
detailed description of the present analysis can be found in~\citep{AugerApJS2012}). 

The Pierre Auger Observatory is located in Malarg\"{u}e, Argentina, at mean latitude 35.2$^\circ\,$S, 
mean longitude 69.5$^\circ\,$W and mean altitude 1400 meters above sea level. It exploits two 
available techniques to detect extensive air showers initiated by UHECRs~: a \textit{Surface Detector (SD) 
array} and a \textit{Fluorescence Detector (FD)}. The SD array consists of 1660 water-Cherenkov detectors 
laid out over about 3000~km$^2$ on a triangular grid with 1.5~km spacing, sensitive to the light 
emitted in their volume by the secondary particles of the showers. At the perimeter of this array, the 
atmosphere is overlooked on dark nights by 27 optical telescopes grouped in 5 buildings.
These telescopes record the number of
secondary charged particles in the air shower as a function of depth in the atmosphere by measuring 
the amount of nitrogen fluorescence caused by those particles along the track of the shower. At the 
lowest energies observed, the angular resolution of the SD is about $2.2^\circ$, and reaches $\sim 1^\circ$ 
at the highest energies. This is sufficient to perform searches for large-scale anisotropies. 
The statistical fluctuation in energy measurement amounts to about 15\%, while the 
absolute energy scale is given by the FD measurements and has a systematic uncertainty 
of 22\%~\citep{AugerPRL2008}. 

In the analyses presented in this letter, the data set consists of events recorded by the SD array from 
1 January 2004 to 31 December 2011, with zenith angles less than 55$^\circ$. To ensure good  
reconstruction, an event is accepted only if all six nearest neighbours of the 
water-Cherenkov detector with the highest signal were operational at the time of the 
event~\citep{AugerNIM2010}. Based on this fiducial cut, any active water-Cherenkov detector with 
six active neighbours defines an active \emph{elemental cell}. In these
conditions, and above the energy at which the detection efficiency saturates, 3~EeV
~\citep{AugerNIM2010}, the total exposure of the SD 
array is 23,520~km$^2$~yr~sr.

 
Due to the steepness of the energy spectrum, any mild bias in the estimate of the shower
energy with time or zenith angle can lead to significant distortions of the event counting rate
above a given energy. It is thus critical to control the energy estimate in searching for anisotropies.
The procedure followed to obtain an unbiased estimate of the shower energy consists in correcting 
measurements of shower signals for the influences of weather effects~\citep{AugerAPP2009} 
and the geomagnetic field~\citep{AugerJCAP2011}. Using the constant intensity cut method~\citep{Hersil1961}, 
the shower signal is then converted to the value that would have been expected had the 
shower arrived at a zenith angle 38$^\circ$. This reference shower signal is finally converted into energy 
using a calibration curve based on hybrid events measured simultaneously by the SD array 
and FD telescopes, since the latter can provide a calorimetric measurement of the 
energy~\citep{AugerPRL2008}. 

In searching for anisotropies, it is also critical to know accurately the effective time-integrated collecting 
area for a flux from each direction of the sky, or in other words, the \textit{directional exposure} $\omega$ 
of the Observatory. For each elemental cell, this is obtained through 
the integration over Local Sidereal Time (LST)
$\alpha^0$ of $x^{(i)}(\alpha^0)\times a_{\mathrm{cell}}{(\theta)}\times\epsilon(\theta,\varphi,E)$, 
with $x^{(i)}(\alpha^0)$ the total operational time of the cell $(i)$ at LST $\alpha^0$, 
$a_{\mathrm{cell}}{(\theta)}=1.95~\cos{\theta}~$km$^2$
the geometric aperture of each elemental cell under incidence zenith angle $\theta$~\citep{AugerNIM2010}, 
and $\epsilon(\theta,\varphi,E)$ the detection efficiency under incidence zenith angle $\theta$ and azimuth
angle $\varphi$ at energy $E$. In the same way as in~\citep{AugerAPP2011}, the small modulation 
of the exposure in $\alpha^0$ due to the variations of $x^{(i)}$ can be accounted for by re-weighting
the events with the number of elemental cells at the LST of each event $k$, $\Delta N_{\mathrm{cell}}(\alpha^0_{k})$. 
Since both $\theta$ and $\varphi$ depend only on the difference $\alpha-\alpha^0$, the integration 
over $\alpha^0$ can then be substituted for an integration over the hour angle $\alpha^\prime=\alpha-\alpha^0$ 
so that the directional exposure actually does not depend on right ascension when the $x^{(i)}$
are assumed to be independent of the LST~:
\begin{equation}
\label{eqn:omega}
\omega(\delta,E) = \sum_{i=1}^{n_{\mathrm{cell}}}x^{(i)}\int_0^{24h}~\mathrm{d}\alpha^\prime\,a_{\mathrm{cell}}{(\theta(\alpha^\prime,\delta))}~\epsilon(\theta(\alpha^\prime,\delta),\varphi(\alpha^\prime,\delta),E).
\end{equation}
The zenithal dependence of the detection efficiency $\epsilon(\theta,\varphi,E)$ can be obtained 
directly from the data in an empirical way~\citep{AugerApJS2012}. Additional effects have an impact 
on $\omega$, such as the azimuthal dependence of the efficiency due to geomagnetic effects,
the corrections to both the geometric aperture of each elemental cell and the detection efficiency
due to the tilt of the array, and the corrections due to the spatial extension of the array. Accounting
for all these effects, the resulting dependence of $\omega$ on declination can be found 
in~\citep{AugerApJS2012}. For a wide range of declinations between $\simeq -89^ \circ$ and
$\simeq -20^ \circ$, the directional exposure is $\simeq 2,500~$km$^ 2$~yr at 1~EeV, and 
$\simeq 3,500~$km$^ 2$~yr for any energy above full efficiency. Then, at higher declinations, 
it smoothly falls to zero, with no exposure above $20^ \circ$ declination.


The detection of significant dipole or quadrupole moments above EeV energies would be of 
considerable interest. Dipole and quadrupole patterns are encoded in the low order $a_{1m}$ and 
$a_{2m}$ coefficients of the multipolar expansion of any angular distribution over the sphere 
$\Phi(\mathbf{n})$~:
\begin{equation}
\label{eqn:ylm}
\Phi(\mathbf{n})=\sum_{\ell\geq0}\sum_{m=-\ell}^{\ell}~a_{\ell m}Y_{\ell m}(\mathbf{n}),
\end{equation}
where $\mathbf{n}$ denotes a unit vector taken in equatorial coordinates. Due to the non-uniform
and incomplete coverage of the sky at the Pierre Auger Observatory, the estimated
coefficients $\overline{a}_{\ell m}$ are determined in a two-step procedure. First, from any
event set with arrival directions $\{\mathbf{n_1},...,\mathbf{n_N}\}$ recorded at LST 
$\{\alpha^0_1,...,\alpha^0_N\}$, the multipolar coefficients of the angular distribution
coupled to the exposure function are estimated through~:
\begin{equation}
\label{eqn:blm-est}
\overline{b}_{\ell m}=\sum_{k=1}^{N} \frac{Y_{\ell m}(\mathbf{n}_k)}{\Delta N_{\mathrm{cell}}(\alpha^0_{k})}.
\end{equation}
$\Delta N_{\mathrm{cell}}(\alpha^0_{k})$ corrects for the slightly 
non-uniform directional exposure in right ascension. Then, assuming that the multipolar expansion of 
the angular distribution $\Phi(\mathbf{n})$ is \textit{bounded} to $\ell_{\mathrm{max}}$, 
the first $b_{\ell m}$ coefficients with $\ell\leq\ell_{\mathrm{max}}$ are related to the non-vanishing 
$a_{\ell m}$ through~:
\begin{equation}
\label{eqn:blm-est}
\overline{b}_{\ell m}=\sum_{\ell^\prime=0}^{\ell_{\mathrm{max}}}\sum_{m^\prime=-\ell^\prime}^{\ell^\prime} [K]_{\ell m}^{\ell^\prime m^\prime} \overline{a}_{\ell^\prime m^\prime},
\end{equation}
where the matrix $K$ is entirely determined by the directional exposure~:
\begin{equation}
\label{eqn:K}
[K]_{\ell m}^{\ell^\prime m^\prime}=\int_{\Delta\Omega} \mathrm{d}\Omega~\omega(\mathbf{n})~Y_{\ell m}(\mathbf{n})~Y_{\ell^\prime m^\prime}(\mathbf{n}).
\end{equation}
Inverting Eqn.~\ref{eqn:blm-est} allows us to recover the underlying $\overline{a}_{\ell m}$, with
a resolution proportional\linebreak 
to $([K^{-1}]_{\ell m}^{\ell m}~\overline{a}_{00})^{0.5}$~\citep{alm}. 
As a consequence of the incomplete coverage of the sky, this resolution deteriorates by a factor larger than 
2 each time $\ell_{\mathrm{max}}$ is incremented by 1. With our present statistics, this prevents the recovery 
of each coefficient with good accuracy as soon as $\ell_{\mathrm{max}}\geq3$, which is why we restrict 
ourselves to dipole and quadrupole searches. 

\begin{figure}[!t]
  \centering					 
  \includegraphics[width=10cm]{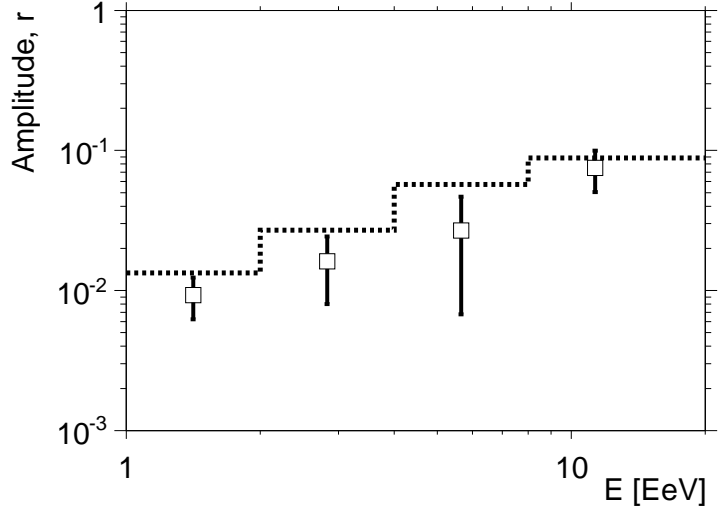}
  \caption{\small{Reconstructed amplitude of the dipole as a function of the energy. The dotted line stands 
  for the 99\% $C.L.$ upper bounds on the amplitudes that would result from fluctuations of an isotropic 
  distribution.}}
\label{fig:ampdip}
\end{figure}
\begin{figure}[!t]
  \centering					 
  \includegraphics[width=10cm]{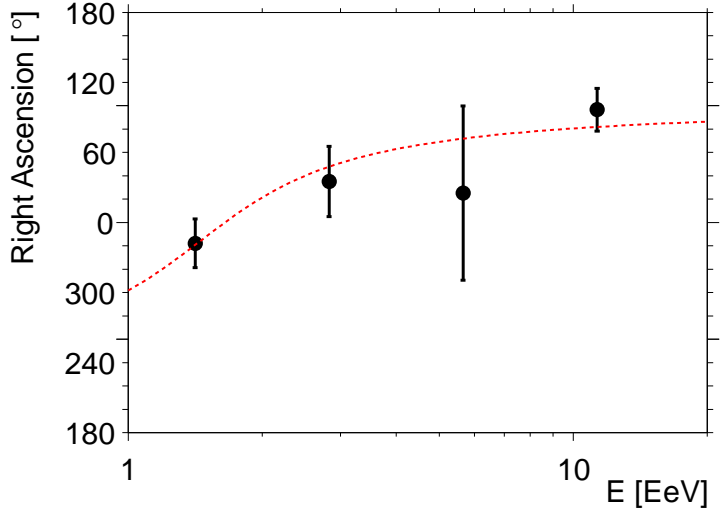}
  \caption{\small{Reconstructed right ascension of the dipole as a function of the energy.
  The smooth fit to the data of~\citep{AugerAPP2011} is shown as the dashed line (see text).}}
\label{fig:radip}
\end{figure}

We first assume that the angular distribution of cosmic rays is modulated by a \emph{pure}  
dipole and parameterise the intensity $\Phi(\mathbf{n})$ in any direction as~:
\begin{equation}
\label{eqn:phi-dip}
\Phi(\mathbf{n})=\frac{\Phi_0}{4\pi}~\bigg(1+r~\mathbf{d}\cdot\mathbf{n} \bigg),
\end{equation}
where $\mathbf{d}$ denotes the dipole unit vector. 
The dipole pattern is here fully characterised by a declination $\delta_d$, a right ascension $\alpha_d$,
and an amplitude $r$ corresponding to the maximal anisotropy contrast~: $r=(\Phi_{\mathrm{max}}-\Phi_{\mathrm{min}})/(\Phi_{\mathrm{max}}+\Phi_{\mathrm{min}})$.
\begin{figure}[!t]
  \centering					 
  \includegraphics[width=9cm]{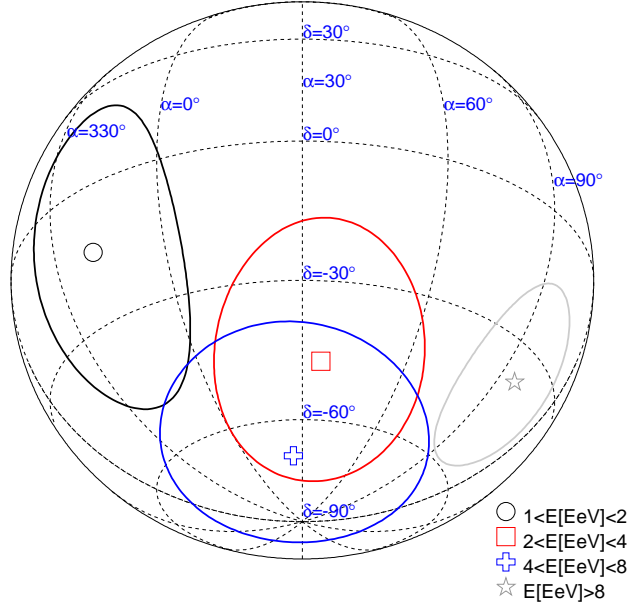}
  \caption{\small{Reconstructed declination and right-ascension  of the dipole with corresponding 
  uncertainties, as a function of the energy, in orthographic projection.}}
\label{fig:dirdip}
\end{figure}
The estimation of these three coefficients is straightforward from the 
estimated spherical harmonic coefficients $\overline{a}_{1m}$. The reconstructed amplitudes 
$\overline{r}$ are shown in Fig.~\ref{fig:ampdip} as a function of the energy. The 99\% $C.L.$ upper 
bounds on the amplitudes that would result from fluctuations of an isotropic distribution are indicated by 
the dotted line. One can see that within the statistical uncertainties, there is no evidence of any 
significant signal. In Fig.~\ref{fig:dirdip}, the corresponding directions are shown in orthographic projection
with the associated uncertainties, as a function of the energy. Both angles are expected to be randomly 
distributed in the case of independent samples 
whose parent distribution is isotropic. It is thus interesting to note that all reconstructed declinations 
are in the equatorial southern hemisphere, and to note also the intriguing smooth alignment of the phases 
in right ascension as a function of the energy. In our previous report on first harmonic analysis in right 
ascension~\citep{AugerAPP2011}, we already pointed out this alignment, and stressed that such a consistency 
of phases in adjacent energy intervals is expected with smaller number of events than the detection of 
amplitudes standing-out significantly above the background noise in the case of a real underlying anisotropy. This 
motivated us to design a \textit{prescription} aimed at establishing at 99\% $C.L.$ whether this consistency 
in phases is real, using the exact same analysis as the one reported in~\citep{AugerAPP2011}. The 
prescribed test will end once the total exposure since 25 June 2011 reaches 21,000~km$^2$~yr~sr.
The smooth fit to the data of~\citep{AugerAPP2011} is shown as a dashed line in 
Fig~\ref{fig:radip}, restricted to the energy range considered here. Though the phase between 
4 and 8~EeV is poorly determined due to the corresponding direction in declination pointing close to the 
equatorial south pole, it is noteworthy that a consistent smooth behaviour is observed using the analysis 
presented here and applied to a data set containing two additional years of data.

\begin{figure}[!t]
  \centering					 
  \includegraphics[width=17cm]{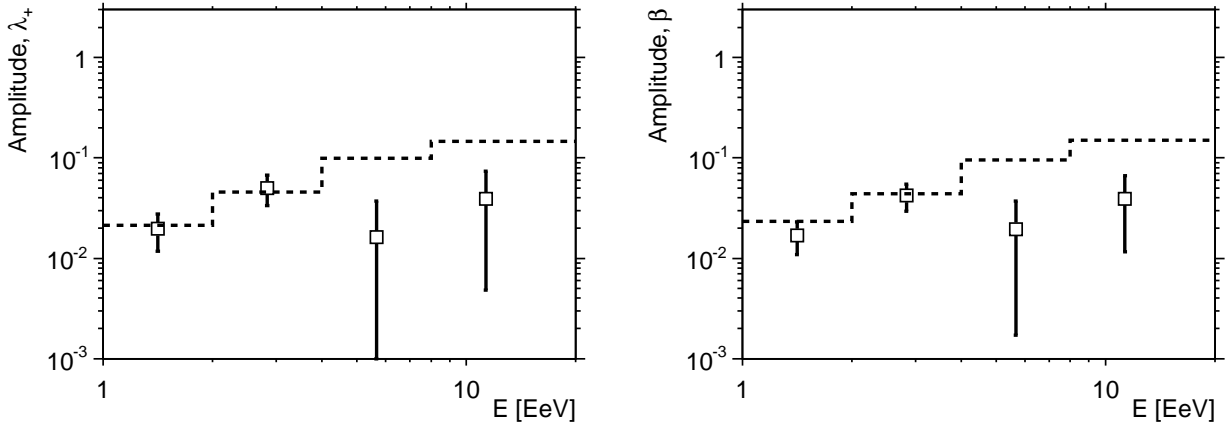}
  \caption{\small{Amplitudes of the quadrupolar moment as a function of the energy using a multipolar 
  reconstruction up to $\ell_{\mathrm{max}}=2$. The dotted lines stand for the 99\% $C.L.$ upper bounds 
  on the amplitudes that could result from fluctuations of an isotropic distribution.}}
\label{fig:ampquad}
\end{figure}

Assuming now that the angular distribution of cosmic rays is modulated by a dipole \emph{and}
a quadrupole, the intensity $\Phi(\mathbf{n})$ can be parameterised in any direction 
$\mathbf{n}$ as~:
\begin{equation}
\label{eqn:phi-quad}
\Phi(\mathbf{n})=\frac{\Phi_0}{4\pi}~\bigg(1+r~\mathbf{d}\cdot\mathbf{n} +\lambda_+(\mathbf{q_+}\cdot\mathbf{n})^2 +\lambda_0(\mathbf{q_0}\cdot\mathbf{n})^2 +\lambda_-(\mathbf{q_-}\cdot\mathbf{n})^2 \bigg),
\end{equation}
with the constraint $\lambda_++\lambda_-+\lambda_0=0$. It is convenient to define the quadrupole 
amplitude $\beta\equiv(\lambda_+-\lambda_-)/(2+\lambda_++\lambda_-)$, which provides a measure of 
the maximal quadrupolar contrast in the absence of a dipole.
Hence, any quadrupolar pattern can be fully described by two amplitudes $(\beta,\lambda_+)$
and three angles~: $(\delta_+,\alpha_+)$ which define the orientation of $\mathbf{q_+}$
and $(\alpha_-)$ which defines the direction of $\mathbf{q_-}$ in the orthogonal plane to 
$\mathbf{q_+}$. The third eigenvector $\mathbf{q_0}$ is orthogonal to $\mathbf{q_+}$ and
$\mathbf{q_-}$. The estimated amplitudes $\overline{\lambda}_+$ and $\overline{\beta}$ are shown 
in Fig.~\ref{fig:ampquad} as functions of the energy. In the same way as for dipole amplitudes, 
the 99\% $C.L.$ upper bounds on the quadrupole amplitude that could result from fluctuations of 
an isotropic distribution are indicated by the dashed lines. Throughout the energy range, 
there is no evidence for anisotropy. 

There are small uncertainties in correcting the estimator of the energy for weather and geomagnetic 
effects, and these propagate into systematic uncertainties in the measured 
anisotropy parameters. As well, anisotropy parameters may be altered in a systematic way by 
energy dependence of the attenuation curve. All these systematic effects have 
been quantified~\citep{AugerApJS2012}. They do not change significantly the results presented here.

\begin{figure}[!t]
  \centering					 
  \includegraphics[width=17cm]{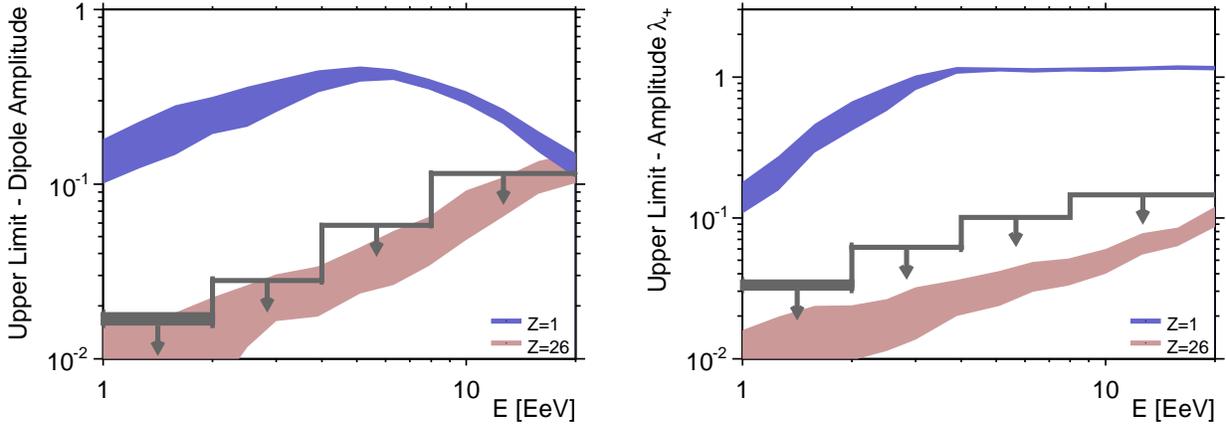}
  \caption{\small{99\% $C.L.$ upper limits on dipole and quadrupole amplitudes as a function of the energy.
  Some generic anisotropy expectations from stationary galactic sources distributed in the disk are
  also shown, for various assumptions on the cosmic ray composition. The fluctuations of the amplitudes
  due to the stochastic nature of the turbulent component of the magnetic field are sampled from different 
  simulation data sets and are shown by the bands.}}
\label{fig:UL}
\end{figure}

From these analyses, upper limits on dipole and quadrupole amplitudes can be derived 
at 99\% $C.L.$. They are shown in Fig.~\ref{fig:UL} for the dipole amplitudes, accounting for the 
systematic uncertainties. We illustrate now their astrophysical interest by calculating the amplitudes 
of anisotropy expected in a toy scenario in which sources of EeV-cosmic rays are stationary, densely 
and uniformly distributed in the galactic disk, and emit particles in all directions.

Both the strength and the structure of the magnetic field in the Galaxy, known only approximately,
play a crucial role in the propagation of cosmic rays. The field is thought to contain a large 
scale regular component and a small scale turbulent one, both having a local strength of a few microgauss 
(see \textit{e.g.}~\citep{Beck2001}). While the turbulent component dominates in strength by a factor 
of a few, the regular component imprints dominant drift motions as soon as the Larmor radius of
cosmic rays is larger than the maximal scale of the turbulences (thought to be in the range 10-100~pc). 
We adopt here a recent parameterisation of the regular component obtained by fitting 
model field geometries to Faraday rotation measures of extragalactic radio sources and polarised
synchrotron emission (BSS-model, with anti-symmetric halo with respect to the galactic 
plane)~\citep{Pshirkov2011}. In addition to the regular component, a turbulent field 
is generated according to a Kolmogorov power spectrum and is pre-computed on a three dimensional 
grid periodically repeated in space. The size of the grid is selected to match the maximal scale of
turbulences (taken here as 100~pc), and the strength of the turbulent component is taken as three 
times the strength of the 
regular one. To describe the propagation of cosmic rays with energies $E\geq 1$~EeV in such 
a magnetic field, the direct integration of trajectories is the most appropriate tool. To obtain 
the anisotropy of cosmic rays emitted from sources uniformly distributed in a cylinder with a radius
of 20~kpc from the galactic centre and with a height of $\pm$ 100~pc, we adopt a method first proposed 
in~\citep{Thielheim1968}. It consists in back-tracking anti-particles with random directions from the 
Earth to outside the Galaxy. Each test particle \textit{probes} the total luminosity along the path 
of propagation from each direction as seen from the Earth. For \textit{stationary sources emitting 
cosmic rays in all directions}, the expected flux in the initial sampled direction is proportional 
to the time spent by each test particle in the source region.

The amplitudes of anisotropy obviously depend on the rigidity $E/Z$ of the cosmic rays, with $Z$ the
electric charge of the particles. Since we only aim at illustrating the upper limits, we consider two 
extreme single primaries~: protons and iron nuclei. The calculation of anisotropy amplitudes for single 
primaries is useful to probe the allowed contribution of each primary as a function of the energy.

The dipole and quadrupole amplitudes obtained for several energy values covering 
the range $1\leq E/\mathrm{EeV}\leq 20$ are shown in Fig.~\ref{fig:UL}. To probe unambiguously 
amplitudes down to the percent level, it is necessary to generate simulated event sets with at least
$\simeq 5~10^5$ test particles. Such a number of simulated events allows us to shrink statistical 
uncertainties on amplitudes at the $0.5$\% level. Meanwhile, there is an intrinsic variance in the model 
for each anisotropy parameter due to the stochastic nature of the turbulent component of the magnetic field. 
This variance is estimated through the simulation of 20 sets of $5~10^5$ test particles, where
the configuration of the turbulent component is frozen in each set. The RMS of the amplitudes sampled 
in this way is shown by the bands in Fig.~\ref{fig:UL}. 

The resulting amplitudes for protons largely stand above the allowed limits. Consequently, unless the 
strength of the magnetic field is much higher than in the picture used here, the upper limits derived 
in this analysis exclude that the light component of cosmic rays comes from galactic stationary sources 
densely distributed in the galactic disk and emitting in all directions. To respect the dipole limits 
below the ankle energy, the fraction of protons should not exceed $\simeq$ 10\% of the cosmic ray composition.
This is particularly interesting in the view of the indications for the presence of a light component
around $1~$EeV from shower depth maximum measurements~\citep{AugerPRL2010,HiResPRL2010,TAAPS2011},
though firm interpretations of these measurements in terms of the atomic mass still suffer from some
ambiguity due to the uncertain hadronic interaction models used to describe the shower developments. 
On the other hand, if the cosmic ray composition around $1~$EeV results from a mixture containing heavy 
elements of galactic origin and light elements of extragalactic origin, upper limits can be respected. This 
is because large scale anisotropy amplitudes below the percent level are expected for extragalactic cosmic 
rays, due to the motion of the Galaxy relative to a possibly stationary extragalactic cosmic ray rest 
frame~\citep{Kachelriess2006,Harari2010}.

Future measurements of composition below $1~$EeV will come from the low energy extension HEAT now 
available at the Pierre Auger Observatory~\citep{AugerHEAT}. Combining these measurements with large 
scale anisotropy ones will then allow us to further understand the origin of cosmic rays at energies 
less than 4~EeV. 

\section*{Acknowledgements}

The successful installation, commissioning, and operation of the Pierre Auger Observatory
would not have been possible without the strong commitment and effort
from the technical and administrative staff in Malarg\"ue.

We are very grateful to the following agencies and organizations for financial support: 
Comisi\'on Nacional de Energ\'ia At\'omica, 
Fundaci\'on Antorchas,
Gobierno De La Provincia de Mendoza, 
Municipalidad de Malarg\"ue,
NDM Holdings and Valle Las Le\~nas, in gratitude for their continuing
cooperation over land access, Argentina; 
the Australian Research Council;
Conselho Nacional de Desenvolvimento Cient\'ifico e Tecnol\'ogico (CNPq),
Financiadora de Estudos e Projetos (FINEP),
Funda\c{c}\~ao de Amparo \`a Pesquisa do Estado de Rio de Janeiro (FAPERJ),
Funda\c{c}\~ao de Amparo \`a Pesquisa do Estado de S\~ao Paulo (FAPESP),
Minist\'erio de Ci\^{e}ncia e Tecnologia (MCT), Brazil;
AVCR AV0Z10100502 and AV0Z10100522, GAAV KJB100100904, MSMT-CR LA08016,
LG11044, MEB111003, MSM0021620859, LA08015 and TACR TA01010517, Czech Republic;
Centre de Calcul IN2P3/CNRS, 
Centre National de la Recherche Scientifique (CNRS),
Conseil R\'egional Ile-de-France,
D\'epartement  Physique Nucl\'eaire et Corpusculaire (PNC-IN2P3/CNRS),
D\'epartement Sciences de l'Univers (SDU-INSU/CNRS), France;
Bundesministerium f\"ur Bildung und Forschung (BMBF),
Deutsche Forschungsgemeinschaft (DFG),
Finanzministerium Baden-W\"urttemberg,
Helmholtz-Gemeinschaft Deutscher Forschungszentren (HGF),
Ministerium f\"ur Wissenschaft und Forschung, Nordrhein-Westfalen,
Ministerium f\"ur Wissenschaft, Forschung und Kunst, Baden-W\"urttemberg, Germany; 
Istituto Nazionale di Fisica Nucleare (INFN),
Ministero dell'Istruzione, dell'Universit\`a e della Ricerca (MIUR), Italy;
Consejo Nacional de Ciencia y Tecnolog\'ia (CONACYT), Mexico;
Ministerie van Onderwijs, Cultuur en Wetenschap,
Nederlandse Organisatie voor Wetenschappelijk Onderzoek (NWO),
Stichting voor Fundamenteel Onderzoek der Materie (FOM), Netherlands;
Ministry of Science and Higher Education,
Grant Nos. N N202 200239 and N N202 207238, Poland;
Portuguese national funds and FEDER funds within COMPETE - Programa Operacional Factores de Competitividade through 
Funda\c{c}\~ao para a Ci\^{e}ncia e a Tecnologia, Portugal;
Romanian Authority for Scientific Reseach, UEFICDI,
Ctr.Nr.1/ASPERA2 ERA-NET, Romania; 
Ministry for Higher Education, Science, and Technology,
Slovenian Research Agency, Slovenia;
Comunidad de Madrid, 
FEDER funds, 
Ministerio de Ciencia e Innovaci\'on and Consolider-Ingenio 2010 (CPAN),
Xunta de Galicia, Spain;
Science and Technology Facilities Council, United Kingdom;
Department of Energy, Contract Nos. DE-AC02-07CH11359, DE-FR02-04ER41300,
National Science Foundation, Grant No. 0450696,
The Grainger Foundation USA; 
NAFOSTED, Vietnam;
Marie Curie-IRSES/EPLANET, European Particle Physics Latin American Network, 
European Union 7th Framework Program, Grant No. PIRSES-2009-GA-246806; 
and UNESCO.

\end{document}